\documentclass[acmsmall,screen]{acmart}

\usepackage{times}
\usepackage{epsfig}
\usepackage{booktabs}
\usepackage{graphicx}
\usepackage{amsmath}
\usepackage{float}
\usepackage{caption}
\usepackage{xcolor}
\usepackage{subfig}
\usepackage{makecell}
\usepackage{ulem}
\usepackage{multirow}
\usepackage{pgf-pie} 
\usepackage{adjustbox} 
\usepackage{chngcntr}
\usepackage{enumitem}
\usepackage[ruled,vlined]{algorithm2e}

\AtBeginDocument{%
  \providecommand\BibTeX{{%
    \normalfont B\kern-0.5em{\scshape i\kern-0.25em b}\kern-0.8em\TeX}}}

\setcopyright{acmcopyright}
\copyrightyear{2022}
\acmYear{2022}
\acmDOI{XXXXXXX.XXXXXXX}

\acmJournal{TOMM}
\acmVolume{1}
\acmNumber{1}
\acmArticle{1}
\acmMonth{3}

\begin{document}

\title {Affective Feedback Synthesis Towards Multimodal Text and Image Data}

\author{Puneet Kumar}
\email{pkumar99@cs.iitr.ac.in}
\orcid{0000-0002-4318-1353}
\affiliation{
	\institution{Indian Institute of Technology Roorkee}
	\country{India}
	\postcode{247667}
}
\author{Gaurav Bhatt}
\email{gauravbhatt@iith.ac.in}
\orcid{0000-0002-0980-8765}
\affiliation{
	\institution{Indian Institute of Technology Hyderabad}
	\country{India}
	\postcode{502285}
}
\author{Omkar Ingle}
\email{iuttam@mt.iitr.ac.in}
\orcid{0000-0001-6310-880X}
\affiliation{
	\institution{Indian Institute of Technology Roorkee}
	\country{India}
	\postcode{247667}
}
\author{Daksh Goyal}
\email{dakshgoyal.171cv111@nitk.edu.in}
\orcid{0000-0002-8999-9142}
\affiliation{
	\institution{National Institute of Technology Karnataka}
	\country{India}
	\postcode{575025}
}
\author{Balasubramanian Raman}
\email{bala@cs.iitr.ac.in}
\orcid{0000-0001-6277-6267}
\affiliation{
	\institution{Indian Institute of Technology Roorkee}
	\country{India}
	\postcode{247667}
}

\renewcommand{\shortauthors}{P. Kumar et al.}

\begin{abstract}
	In this paper, we have defined a novel task of affective feedback synthesis that deals with generating feedback for input text \& corresponding image in a similar way as humans respond towards the multimodal data. A feedback synthesis system has been proposed and trained using ground-truth human comments along with image-text input. We have also constructed a large-scale dataset consisting of image, text, Twitter user comments, and the number of likes for the comments by crawling the news articles through Twitter feeds. The proposed system extracts textual features using a transformer-based textual encoder while the visual features have been extracted using a Faster region-based convolutional neural networks model. The textual and visual features have been concatenated to construct the multimodal features using which the decoder synthesizes the feedback. We have compared the results of the proposed system with the baseline models using quantitative and qualitative measures. The generated feedbacks have been analyzed using automatic and human evaluation. They have been found to be semantically similar to the ground-truth comments and relevant to the given text-image input.  
\end{abstract}

\ccsdesc[500]{Computing methodologies~Natural language generation}
\ccsdesc[300]{Computing methodologies~Machine learning}
\ccsdesc[300]{Information systems~Multimedia and multimodal retrieval}
\ccsdesc[300]{Information systems~Sentiment analysis}

\keywords{Affective Computing, Feedback Synthesis, Multimodal Input, Dataset Construction, Context Vector.}

\maketitle

\section{Introduction}\label{sec:intro}
Multimodal data processing has emerged as an important sub-domain in Artificial Intelligence (AI) research due to the fast growth of multimedia data in the last few years \cite{li2016multimedia}. One of the goals of AI is to enable machines to respond to multimodal data just like humans do. A response could be feedback, answer, caption, vocal signal, facial reaction, bodily gesture, etc.~\cite{page2019multimodal}. Humans portray different emotions through various modalities, among which text and image are known to contain the human emotions and intentions most effectively~\cite{makiuchi2021multimodal}. Machines and systems capable of generating affective feedback towards multimodal text and image data could be very useful~\cite{gallo2020predicting}. Here the term `affective' or `human-like' is used in the sense that the feedback synthesis system should be able to synthesize feedback towards a multimodal input in a way humans do. 

\begin{figure}[!t]
	\centering
	\includegraphics[width=0.95\textwidth]{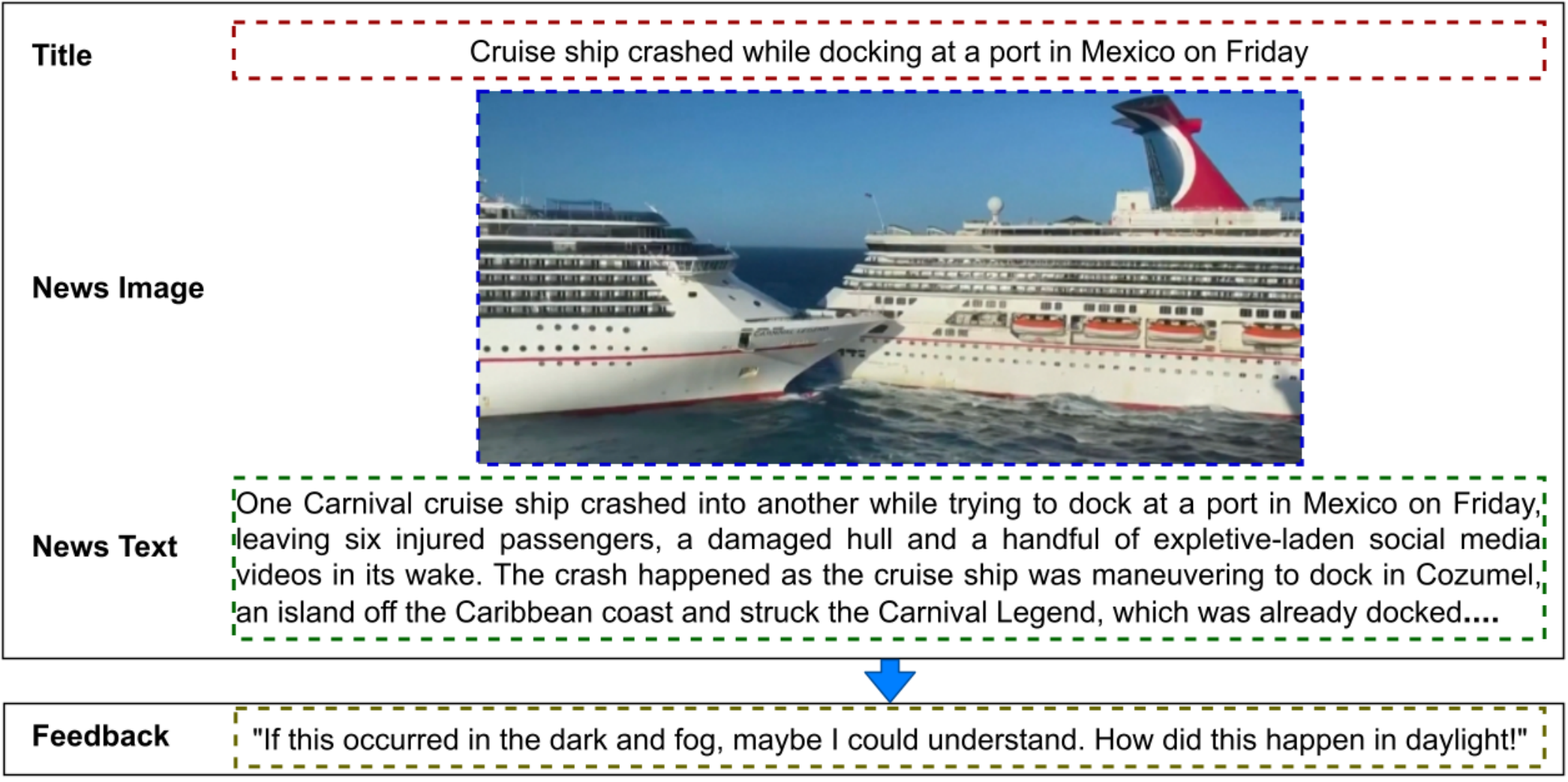}
	\caption{Illustration of the proposed task. It aims to automatically generate feedback (response, reaction, or summary) considering the input's visual \& textual context in a similar way humans do it.}
	\label{fig:intro}
\end{figure}

The ability to synthesize feedback towards multimodal data could be helpful in various applications such as determining user response towards products, social behavior analysis, evaluating multimodal educational content (for example, slides, blogs \& books), and predicting the user engagement in advertisements~\cite{gallo2020predicting}. Multimodal feedback synthesis systems can also be used to predict emotions that multimedia contents would induce in the users and hence predict the success of the contents \cite{muszynski2019recognizing}. Likewise, the success of advertisements and educational content can also be predicted upfront by gauging the kind of feedback users are likely to have towards them \cite{blikstein2016multimodal}. Moreover, the analysis of multimodal data generated by patients can help predict their mental-states \cite{mccutcheon2021increasing}.

Among the existing research tasks, multimodal summarization and dialog generation are related to the affective feedback synthesis task introduced in our work~\cite{gao2020standard,gu2019dialogwae,kang2019dual}. The text and image summarization approaches use the information from a single modality. Whereas multimodal summarization leverages the information from multiple modalities to fine-tune the summary \cite{zhu2018msmo}. However, most multimodal summarization approaches do not use actual human responses to train their models. Likewise, textual and visual dialog generation models are also not trained on human responses. They also suffer from the problem of short and uninformative dialogs' generation \cite{wu2019response}. Moreover, affective feedback generation can be abstract and more difficult compared to the aforementioned tasks. For instance, identifying $sarcasm$ is a common challenge in affective feedback generation that can alter the conveyed sentiment. To this end, the proposed multimodal feedback synthesis system has been trained using actual human comments along with text and image inputs. The relevance of the comments is decided based on their number of likes (upvotes). We have incorporated textual and visual features that enable the proposed system to generate informative feedback considering the textual and visual context of the inputs.

In this paper, we have proposed a novel task to synthesize affective feedback towards text-image inputs, which is illustrated in Fig. \ref{fig:intro}. The idea is to build a model that could generate contextually relevant feedback as a new modality from two given modalities, i.e., news image and text. Affective feedback can be considered a special type of summary produced by humans based on their state of mind induced on processing the emotional context of the inputs \cite{mcdarby2003affective}. Affective feedback aligns with the sentiment being expressed and is potentially useful in analyzing human behavior. We have also constructed a large-scale dataset, i.e., IIT Roorkee Multimodal Feedback (IIT-R MMFeed) dataset, for multimodal feedback synthesis. For simplicity, the `IIT-R MMFeed' dataset has been referred to as the MMFeed dataset in this paper. The MMFeed dataset has been constructed by crawling the news articles through Twitter feeds, and it contains images, text, human comments, and the number of likes for the comments.

The proposed system contains textual and visual encoders (as shown in Fig.~\ref{fig:methodology}) to which input text and image are given as the inputs. The textual encoder uses a multi-headed self-attention-based transformer~\cite{vaswani2017attention} and extracts textual features in the form of textual context vectors. The visual encoder uses a Faster Region-Based Convolutional Neural Networks (R-CNN) model~\cite{ren2015Faster} and extracts visual features in the form of a visual context vector. Each text encoder output vector is concatenated with a visual context vector and passed through a feedforward network that outputs multimodal contextual vectors with the same dimension as the text encoder vectors. Further, the decoder block takes textual context vector, multimodal context vectors, and ground-truth comments as inputs and generates the feedback as output. The generated feedbacks are evaluated using qualitative and quantitative methods for their relevance with the ground-truth comments and inputs. \vspace{.05in}

\noindent The important contributions made in this paper are listed as follows:\vspace{-.05in}
\begin{itemize}
	\item 
	The novel task has been defined to generate feedback towards multimodal input containing image and text similar to how humans do. The task aims to synthesize a new modality, i.e., feedback that is contextually relevant to the two given modalities, i.e., news image and text.
	
	\item 
	A large-scale dataset, `IIT-R MMFeed dataset,' has been constructed for multimodal feedback synthesis by crawling news articles through Twitter feeds. It contains images, text, and Twitter user comments, along with the number of likes (upvotes) for each comment. 
	
	\item 
	A multimodal feedback synthesis system has been proposed. The textual and visual encoders extract textual and visual features using a text transformer and a Faster R-CNN model. The decoder synthesizes the feedback according to the multimodal context that combines textual and visual features. 
	
\end{itemize}


The organization of the remainder of the paper is as follows: the research tasks related to affective feedback synthesis are surveyed in Section~\ref{sec:lr}. Section~\ref{sec:proposed} describes the dataset construction process and the proposed system. The experiments and results have been discussed in Section~\ref{sec:experiments} and Section~\ref{sec:results} respectively. The major conclusions are drawn in Section~\ref{sec:conclusion} along with the future research directions.

\section{Related Work}\label{sec:lr}
This Section surveys the research advances related to the task of affective feedback synthesis introduced in this paper.


\subsection{Text Summarization}
There are two major types of text summarization approaches, viz. abstractive and extractive summarization~\cite{el2021automatic,gao2020standard}. The extractive summarization is focused on extracting useful words from the given text. In this context, Narayan et al. \cite{narayan2017neural} worked on neural extractive summarization and utilized side information such as the title and captions of the news images. The abstractive summarization expresses the main content of a given text using different words instead of creating the summary by selecting the words or sentences from the text. The abstractive summarization of the sentences has been explored through neural attention-based models \cite{rush2017neural}. The extractive summarization is limited by the vocabulary set of the input text, whereas abstractive summarization approaches are not able to generate factually consistent and human-like summaries \cite{kryscinski2020evaluating}. Moreover, text summarization does not leverage the information from multiple modalities to fine-tune the summary. 

\subsection{Image Summarization}
Summarization of a scene from an image collection is an existing research problem \cite{simon2007scene}. In this context, Pan et al. \cite{pan2019content} proposed an approach to generate a visual summarization for a set of images. They considered the social attractiveness features such as image quality and aesthetics while summarizing the images. On the other hand, Samani et al. \cite{samani2018multi} considered semantic features along with social attractiveness features to summarize the images in a context-sensitive manner. The image summarization approaches consider only visual information to produce the summary of the contents, whereas the multimedia contents often contain a combination of visual and textual information~\cite{lahat2015multimodal}. Moreover, the output of these approaches is in visual form only, which is more challenging to infer the emotion and context-related information as compared to textual and multimodal outputs \cite{kumar2021hybrid}. The aforementioned limitation of text summarization and image summarization of not leveraging the information from multiple modalities are addressed by multimodal summarization approaches~\cite{zhu2018msmo}.

\subsection{Multimodal Summarization}
In the context of multimodal summarization, Chen and Zhuge \cite{chen2018abstractive} used recurrent networks to generate a summary from text and images. They performed simultaneous summarization of images and text documents and aligned the sentences and images to generate the summaries. On the other hand, Zhu et al.~\cite{zhu2018msmo} used a pointer generator network for multimodal summarization instead of performing manual alignment of the text and images. They picked the most relevant image among the input images for a data sample and selected the essential keywords from the text inputs by performing extractive summarization. Summarization of multimedia news has also been explored using extractive summarization approaches \cite{chen2015extractive}. In an attempt to utilize videos instead of images for multimedia news articles, Li et al.~\cite{li2020vmsmo} implemented self-attention to automatically choose a suitable video frame based on the semantic meaning of the article. They modeled the article's semantic meaning along with the input video jointly.

Multimodal processing faces the challenges of either missing out on specific modalities or biasing the results on a particular modality~\cite{page2019multimodal}. Zhu et al. used Multimodal reference~\cite{zhu2020multimodal} to handle the modality-bias problem in multimodal summarization. They used the multimodal reference for guidance, designed an objective function, and proposed a novel evaluation metric based on the joint multimodal representation that considered the loss of image selection and summary generation. Context-aware summarization techniques have also been explored for multimodal summarization. In this direction, Li et al.~\cite{li2020aspect} incorporated aspect coverage and corresponding use-cases for various product categories. Synthesis of affect during multimodal summarization has not been explored to its full potential. Moreover, the abovementioned research did not consider human comments towards the multimodal input data while training their summarization models. On the other hand, the proposed system is trained on human-generated comments along with multimodal inputs containing text and images. It generates the feedback as a new modality from two input modalities, i.e., image and text. 


\subsection{Textual Dialog}
Textual dialog systems consider text input questions and generate text output responses in the form of a dialog. In this context, Zhou et al. \cite{zhou2018emotional} developed chatting agents that maintain and use a memory of emotional keywords. In another work, Xu et al.~\cite{xu2018better} worked on conversation modelings for dialog generation, whereas \cite{zhao2017learning} used conditional variational autoencoders to understand the diverse use-cases for dialog generation models. Researchers have also worked on increasing the relevance, diversity, and originality of generation results \cite{xinnuo2018better}. In this context, Gu et al.~\cite{gu2019dialogwae} trained a conditional generative network for dialogue modeling. They modeled the data distribution by training a generative adversarial network considering multiple possible probability distributions of the topics and sentiments in the latent variable space. In an attempt to incorporate human-ness in the generated dialogues, Zhao et al. \cite{zhou2018mojitalk} used conditional autoencoders and included emojis in the synthesized responses. Most of the real-life multimedia context is expressed through multiple modalities where combining the complementary information from various modalities helps understanding the underlying emotional context effectively \cite{hu2018multimodal}. However, textual dialog systems consider only the textual context while generating the response towards input questions. There is a need to consider the corresponding visual context as well, which has given rise to the development of visual dialog systems as explored in the following Section.

\subsection{Visual Dialog}
Visual dialog (VisDial) is another closely related task to Multimodal Affective Feedback Synthesis, which aims to generate dialogs about visual input contents. In this context, Kang et al.~\cite{kang2019dual} proposed a Dual Attention Network (DAN) to resolve visual reference between given image and dialogue history. They implemented multi-head attention to learn the relationships between the given question and the dialog history and bottom-up attention to model the image features and output dialogs' representations. In a similar work, Chen et al.~\cite{chen2020dmrm} implemented dual-channel reasoning to learn the context from the image and dialog history together. The dual-channel reasoning proposed by them enabled them to learn rich semantic representations of the questions compared to the single-channel reasoning approach followed by DAN. In another work, Niu et al. \cite{niu2019recursive} used recursive attention for finding the image component, referred to by a particular text entity. Their work was extended by Park et al. \cite{park2021multi} for multi-view settings of recursive attention. Multimodal VisDial synthesis has also been explored for audio-visual data \cite{alamri2019audio,hori2019end}. To effectively model the visual features, Jiang et al. \cite{jiang2020dualvd} implemented a region-based graph attention network to learn question-aware relationships of the input image and dialog history-aware question features. This inspired us to incorporate attention-based visual feature extraction in our proposed work. 

The existing VisDial methods are not able to solve our problem of affective feedback synthesis. We have empirically found that they could not synthesize meaningful feedback, and the responses are limited to a few words. A little work has been carried out in this context. For instance, Wu et al. \cite{wu2019response} attempted to handle the problem of short and uninformative dialogs generation by editing the generated dialog with informative words based on the context of the input. We have incorporated textual and visual features that enable the proposed system to generate feedback considering the textual and visual context of the inputs. The appropriate textual and visual feature extraction networks have been determined through extensive ablation studies in Section \ref{sec:ablation}. GRU, Bidirectional GRU, and Transformer have been explored for textual feature extraction. The VGG and ResNet models have been explored to extract global visual features, whereas R-CNN variants have been explored for local feature extraction. Moreover, the attention mechanism has been incorporated in textual and visual encoders.

\subsection{Evaluation of Machine Generated Sentences}
The quality of machine-generated sentences such as translations, text summaries, image captions, dialogues, etc., is subjective. Various quantitative metrics have been used in the literature for their evaluation. The importance of their extensive evaluation has been pointed out \cite{bhandari2020re}. Various quantitative metrics based on recall, precision, and sensitivity have also been used. For example, Zhu et al. \cite{zhu2020multimodal} used a recall-oriented metric, ROUGE, to evaluate the machine-generated visual dialog against the textual reference sentence. They also evaluated their output against the reference description of input images using precision value. The ranking-based evaluation metrics such as `Recall@k'~\cite{runeson2007detection} and `Mean Reciprocal Rank'~\cite{craswell2009mean} have also been utilized to rank the machine-generated sentences against a set of references sentences.

Further, multimodal summarization and VisDial systems are prone to the modality-bias problem where they tend to consider one of the input modalities more than the others \cite{zhu2020multimodal}. Subjective human evaluation is helpful to evaluate whether the output is relevant to both the input modalities, i.e., text and image \cite{ladhak2020wikilingua}. For the exhaustive evaluation of the feedback generated by the proposed system, we have implemented five metrics based on recall, precision, and precision (namely BLEU Score, ROUGE, Meteor, CIDEr, and SPICE) and two ranking based metrics (Recall@k and Mean Reciprocal Rank). Human evaluation has also been carried out to subjectively evaluate the relevance of the generated feedback with the input text, image, and ground-truth comments.    

As discovered from the literature surveyed above, synthesis of affect during multimodal summarization has not been explored to its full potential. Most of the above methods generated a textual response towards multimodal data; however, they did not use actual human responses to train their models. An adequate dataset for multimodal feedback synthesis is also not available. With that as an inspiration, a method to synthesize affective feedback towards multimodal data has been proposed in this work. A large-scale dataset has also been constructed for multimodal feedback synthesis.



\section{Proposed system}\label{sec:proposed}
This paper proposes a multimodal system to synthesize feedback for given text and image data as humans do. The formulation of the proposed task has been described as follows, along with the construction of the MMFeed dataset and proposed affective feedback synthesis system's architecture. 

\subsection{Problem Formulation} 
Given a multimodal input $M$ = $\{T, I\}$, where $T$ = $\{t_1, t_2, \dots t_m\}$ is the news text and $I$ is an image ($m$ denotes the length of the news text sequence), the proposed system generates affective, i.e., human-like feedback $F$ = $\{f_1, f_2, \dots f_k\}$ where $k$ denotes the length of the feedback sequence. The problem is to generate a new modality, i.e., feedback from two given modalities, i.e., news image and text. The generated feedback is considered `affective' or `human-like' in the sense that the proposed system synthesizes them in a way humans do. 


\subsection{Dataset Construction}\label{sec:dataset}
A large-scale dataset, IIT Roorkee Multimodal Feedback (IIT-R MMFeed) dataset, has been constructed for multimodal feedback synthesis. The procedure to compile and pre-process the dataset has been described in Appendix \ref{appendix1}. It has been collected by crawling news articles from corresponding Twitter handles (such as `TIME,' `CNN,' `NYTimes,' `BBCBreaking,' etc.) using NLTK\footnote{https://nltk.org/} \& newspaper3k\footnote{https://newspaper.readthedocs.io/} libraries and Tweepy API\footnote{https://docs.tweepy.org/en/stable/}. The MMFeed dataset consists of 77,790 samples collected through 9,479 Tweets containing image, text, and user comments, along with the number of likes (upvotes) for each comment. Table \ref{tab:data} describes various parameters of the dataset while example data instances and the dataset's distribution as per the number of comments each image has been depicted in Table \ref{tab:sample_data} and Fig. \ref{fig:data} respectively. 

The MMFeed dataset stands out from the existing datasets such as `New York Times Articles \& Comments (2020)' dataset \cite{nytdataset2020} as it contains the images as well as the no. of likes (upvotes) that are helpful to evaluate the relevance of the comments in the context of the input image \& text. The MMFeed dataset includes data from various genres such as sports, politics, current affairs, etc., which can be utilized for the proposed feedback synthesis system's robust training. As the human users have generated comments in response to input text and image, they can be considered the combined representation of the textual and visual contexts. Moreover, the number of likes associated with the comments can be considered to denote the comments' relevance to the input image and text. The proposed system has been trained on the input text and image and evaluated against the comments.\vspace{.2in} 

\begin{minipage}{.95\textwidth}
	\begin{center}
		\centering
		\begin{minipage}[!h]{0.44\textwidth}
			\captionof{table}{\centering MMFeed dataset's parameters and their values\vspace{-.1in}}
			\label{tab:data}
			{\fontsize{9}{11}\selectfont
				\resizebox{1\textwidth}{!}
				{
					\begin{tabular}{@{}lc@{}}
						\toprule
						\textbf{Parameter}               & \textbf{Value} \\ \midrule
						No. of news articles             & 9,479          \\
						No. of samples                   & 77,790         \\
						Avg. comments per article        & 8.21           \\
						Avg. no. of likes per comment    & 1.51           \\
						Avg. length of news text         & 611 words      \\
						Avg. length of comments          & 15.71 words    \\ \bottomrule
					\end{tabular}%
				}
			} 
		\end{minipage}
		\hfill \hspace{.3in}
		\begin{minipage}[!h]{0.44\textwidth}
			\centering
			\begin{tikzpicture}[scale=0.43]
				\pie{1/>100c, 3/50-99c, 7/20-49c, 11/13-20c, 24/5-12c, 30/3-5c, 24/1-2c}
			\end{tikzpicture}
			\vspace{-.05in}
			\captionof{figure}{\centering MMFeed dataset's distribution as per no. of comments (denoted as `c').}
			\label{fig:data}
			\vspace{-.05in}
		\end{minipage}
	\end{center}
\end{minipage} 
\vspace{.2in}

\begin{table}[!h]
	\centering	
	\caption{A few samples from the MMFeed dataset. Here, `likes' are the number of likes (upvotes) for the corresponding comment denoting its relevance to the input text and image.\vspace{-.05in}}
	\label{tab:sample_data}
	\resizebox{1\textwidth}{!}{%
		\begin{tabular}{l}
			\includegraphics[width=1\textwidth]{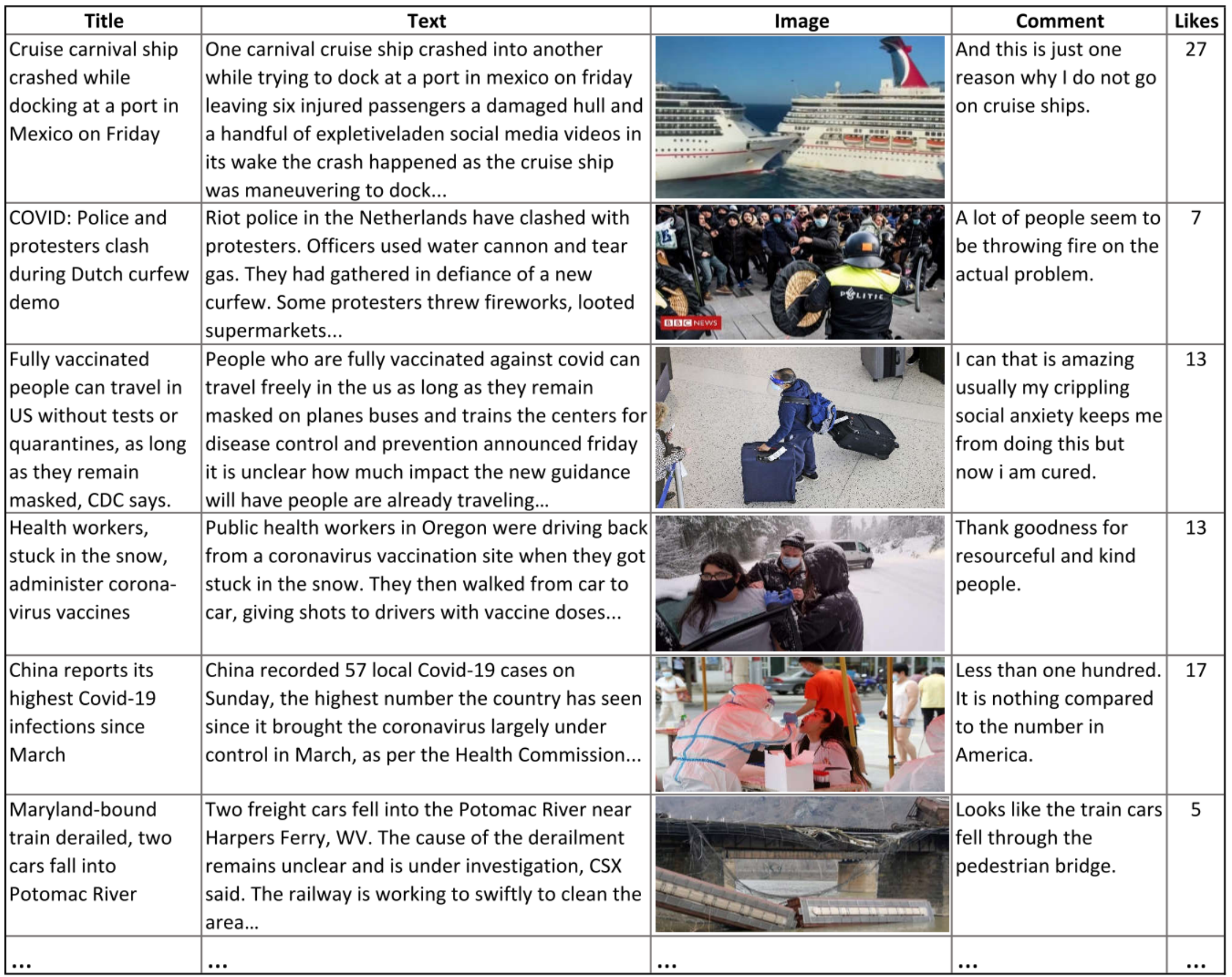}
		\end{tabular}%
	} 
\end{table}




\subsection{Proposed System}
The proposed system's overall architecture is described in Fig.~\ref{fig:methodology} and elaborated in the following Sections. The first two blocks are textual and visual encoders that work in parallel. Input text and image are given as the input to these blocks, respectively. The output of the textual encoder is a time series of vectors (textual context vectors, z*), and the output of the visual encoder is a single vector (visual context vector, g*). Each text encoder output vector is concatenated with g* and passed through a feedforward network that outputs another time series of vectors (multimodal contextual vectors, y*) with the same dimension as the text encoder vectors. Next is the decoder block for which z* and y* are the two inputs. During training, the ground-truth comments are provided as the third input against which the model is trained. During testing, feedback is generated as an output. The similarity module evaluates the similarity between the input comments and generated feedback. 

\begin{figure}[!t]
	\centering
	\includegraphics[width=1\textwidth]{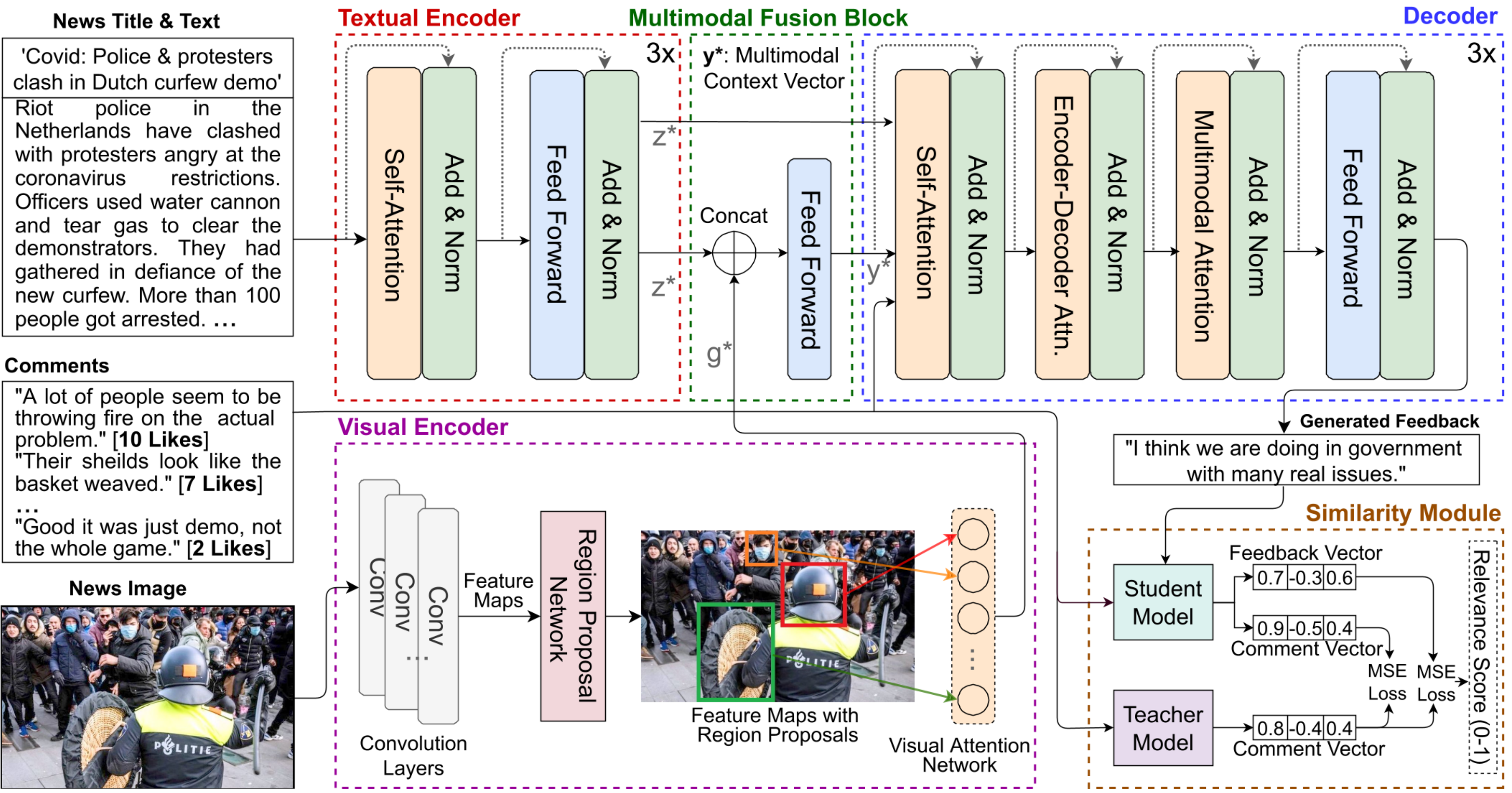}
	\caption{The schematic architecture of the proposed system. Here, three textual encoder blocks and three decoder blocks are used along with a visual encoder network. The dotted arrows represent the residual connections. The similarity module generates the similarity score, which denotes the significance of the generated feedback.}
	\label{fig:methodology}	
\end{figure}




\subsubsection{Text Encoder}
Encoding of textual data is achieved by multi-headed self-attention-based transformer ~\cite{vaswani2017attention}. The encoder block includes a self-attention layer and a feedforward layer. Its sub-layers (i.e., feedforward and self-attention layers) have residual connections around them, and each layer is followed by a normalization layer.\vspace{.08in}

\noindent \textit{Textual Attention}:
Three vectors, $Q$: query, $K$: key, $V$: value are obtained by multiplying encoder's each input with three weight matrices $W(Q), W(K), W(V)$ and used to find the attention head $z$ are per Eq. \ref{eq:visatn}.
The attention heads and key vector's dimension are denoted by $\{z_1, z_2, ... z_n\}$ and $d_k$. 
{\fontsize{9}{11}\selectfont
	\begin{eqnarray}
		\label{eq:visatn}
		\begin{split}	
			&\ \ \ z = Attention(Q, K, V ) = softmax(\frac{Q.K^T}{\sqrt{d_k}})V\\ 	
		\end{split}
	\end{eqnarray}
} 
\noindent Where $Q$, $K$, and $V$ are the matrices packed with all the queries, keys, and values, respectively, whereas $K^T$ denotes the Transpose of the matrix $K$ and $d_k$ is the scaling factor.

We concatenated the attention heads and multiplied the weight matrix $W^0$ to generate an intermediate vector $z'$, which is passed through the feed-forward layer to generate the textual context vector, $z^*$. The multiple attention layers are stacked to learn the information simultaneously from different representation positions. The keys, values and queries are linearly projected $h$ times to $d_k$, $d_v$ and $d_k$ dimensions. As shown in Eq.~\ref{eq:mha}, the attention function is operated on the projected keys, values, and queries simultaneously to get $d_v$-dimensional output. 

{\fontsize{9}{11}\selectfont
	\begin{eqnarray}
		\label{eq:mha}
		\begin{split}	
			&\ \ \ MultiHead(Q, K, V) = Concat(head_1, ..., head_h)W^O\\
			&\ \ \ head_i = Attention(QW^{Q}_{i}, KW^{K}_{i}, VW^{V}_{i})\\	
		\end{split}
	\end{eqnarray}
}

\noindent Where $h$ denotes the number of attention heads whereas $W^{Q}_{i}$, $W^{K}_{i}$, $W^{V}_{i}$ and $W^{O}_{i}$ are the parameter matrices with the projections of queries, keys, values, and output, respectively.

The encoder stack has a fully connected feed-forward network (FFN) for each layer. The input layer of this network is of 256 (Need to verify) dimensions, which is the same as the dimension of the output layer, while the hidden layers are of 512 dimensions. The network's output for input $x$ is computed as per Eq.~\ref{eq:ffn}.

{\fontsize{9}{11}\selectfont
	\begin{eqnarray}
		\label{eq:ffn}
		\begin{split}	
			&FFN(x) = max(0, xW_1 + b_1)W_2 + b_2 \\ 	
		\end{split}
	\end{eqnarray}
}

\noindent Where $x$ is the input; $b_1$ \& $b_2$ are bias terms, and $W_1$ \& $W_2$ are the weight matrices.


\subsubsection{Image Encoder} 
We have used a pre-trained Faster R-CNN~\cite{ren2015Faster} model to extract the visual features. The input image is fed to a series of convolutional layers to obtain the feature maps. Then the RPN runs a sliding window of size $n*n$ through them to generate the anchor boxes. The total number of anchor boxes for a feature map is $W_i*H_i*K_i$, where $H_i$ and $W_i$ denote the height and width of a feature map, and $K_i$ is the number of anchors for each position. An objectiveness score defined as the Intersection Over Union (IoU) is given to each anchor box. This score can be either positive or negative (Eq. \ref{eq:rpnos}). The anchors which do not have a score associated with them do not contribute to the training. These anchors are passed into the classification (object classification) and regression (object localization) layers which output the classified boxes in the image. 

{\fontsize{9}{11}\selectfont
	\begin{eqnarray}
		\label{eq:rpnos}
		Objectiveness\ Score_{(IoU)} = \begin{cases}
			Positive; \hspace{.4in} IoU > 0.7 \\ 
			Positive; \hspace{.4in} 0.5 \leq IoU < 0.7 \\ 
			Negative; \hspace{.35in} IoU < 0.3 \\ 
			Not\ Negative; \hspace{.12in} 0.3 \leq IoU \leq 0.5 \\ 
		\end{cases}
	\end{eqnarray}
} 

Eq. \ref{eq:rpnlf} shows the loss function used for training the image encoder model. Here, 1601 classes were used for the anchor box. The model is trained to classify the anchor box, and its output is passed to the Visual Attention Network.  

{\fontsize{9}{11}\selectfont
	\begin{eqnarray}
		\label{eq:rpnlf}
		\begin{split}
			& L(\{p_i\}, \{t_i\}) = \frac{1}{N_{cls}}\sum_{i}L_{cls}(p_i, p^*_i) + \lambda\frac{1}{N_{reg}}\sum_{i}p^*_{i}L_{reg}(t_i, t^*_i) \\
		\end{split}
	\end{eqnarray}
}

\noindent Where $p_{i}$ \& $p^{*}_{i}$ are the predicted probability and ground-truth value of anchors denoting whether they contain an object or not; $t_{i}$ denotes the coordinates of the predicted anchors; $t^{*}_{i}$ denotes the ground-truth coordinates associated with the bounding boxes; $L_{cls}$ \& $L_{reg}$ are the classifier loss and regression loss terms; $N_{cls}$ \& $N_{reg}$ denote the normalization parameters for mini-batch (of size 256) and regression and $\lambda$ is a hyper-parameter with a value of 10. 

The rationale behind choosing a Faster R-CNN for object detection during visual feature extraction is governed by its speed and applicability in our use case. The Faster R-CNN model is computationally Faster in extracting the visual features than the other models of its family (namely, R-CNN \cite{girshick2014rich}, and Fast R-CNN \cite{girshick2015Fast} that incorporated ROI (Region of Interest) pooling layer to speed up the conventional R-CNN model). The Faster R-CNN replaces the conventional selective search-based region proposal with a Region Proposal Network (RPN). The reason why Faster R-CNN was used over the much Faster You only look once (YOLO) model~\cite{redmon2016you} is due to the inefficiency of YOLO while identifying small objects. We could not afford to lose any such information in our image data. The YOLO model also struggles in identifying objects with skewed aspect ratios, while this is not the case with the Faster R-CNN model. Moreover, the proposed task of feedback synthesis does not need to be real-time. Finally, the Faster R-CNN model was chosen for the implementation.\vspace{.08in}

\noindent \textit{Visual Attention}: Visual context vector $g^*$ is computed using Eq.~\ref{eq:textatn} where $g$ is the global feature vector while $c_i^{s}$, $b_i^{s}$, $a_i^{s}$ denote compatibility score, feature vector and attention vector for state $s$ and anchor box $i$.\vspace{-.15in}  

\begin{eqnarray}
	\label{eq:textatn}
	\begin{split}		
		&\ c_i^{s} = (b_i^{s})^T.g; \\  
		&a_i^{s} = \frac{exp(c_i^{s})}{\sum_{j}^{n}exp(c_j^{s})};\\  
		&g^{*} = \sum_{i=1}^{n}(a_i^{s} \cdot b_i^{s})\\ 
	\end{split}
\end{eqnarray}

\noindent Where $c_i^{s}$ denotes the compatibility score for state $s$ and box $i$; $g$ is the global feature vector; $g^*$ is the visual context vector (final output of attention for state $s$) whereas $b_i^{s}$ \& $a_i^{s}$ denote the feature vector \& attention vector for state $s$ and box $i$. 


\subsubsection{Multimodal Fusion Block} 
Generation of multimodal context is achieved by combining textual context and visual context in Multimodal fusion block. The output of the text encoder (textual context vectors,$z^*$) and the output of the visual encoder (visual context vector, $g^*$) are given as inputs to the Multimodal fusion block. The Multimodal Fusion block computes the multimodal context vectors as follows – The output of the text encoder is a time series of vectors. Each of these vectors is concatenated with g* and passed through a feedforward network that outputs a time series of vectors with the same dimension as the text encoder vectors, which is termed as Multimodal context vectors ($y^*$). The attention module incorporates textual, visual, and multimodal attention. The textual and visual attention have been described above, whereas multimodal attention is as follows.\vspace{.08in}

\noindent \textit{Multimodal Attention}: 
Multimodal context vector $y^*$ is calculated in Eq.~\ref{eq:multiatn1} by concatenating $z^*$ \& $g^*$ and passing through a feed-forward layer. Further, $z^*$ and $y^*$ are fed as input to each decoder block. \vspace{-.1in} 

{
	\begin{eqnarray}
		\label{eq:multiatn1}
		\begin{split}		
			&y^*=concat(z^*,g^*)^T.W\\ 
		\end{split}
	\end{eqnarray}
} 
\noindent Where $y^*$, $z^*$ and $g^*$ denote multimodal, textual, and visual context vectors; $W$ is the weight matrix, and $T$ denotes the transpose operation.


\subsubsection{Decoder} The decoder block takes multimodal context vectors and $z^*$, textual context vectors, $y^*$ as input. During training, it takes the ground-truth comments as additional input against which the model is trained. The comments' embeddings are produced, and positional encoding is added for them in a similar way to how the encoder encodes the text. During testing, the decoder synthesizes the feedback using $z^*$ and $y^*$. Various layers of the decoder block are listed as follows. Each attention layer has a residual connection around it, and a normalization layer follows it.
\begin{enumerate} [label=(\alph*)]
	\item 
	\textit{Self-attention layer}: where query, keys, and values are the decoder representation.\vspace{.05in}
	\item  
	\textit{Encoder-decoder attention layer}: where a query is decoder representation and $z^*$ is assigned to keys and values.\vspace{.05in}
	\item  
	\textit{Multimodal-attention layer}: implemented to achieve multimodal attention with architecture similar to self-attention layer except the queries are decoder representations and $y^*$ is assigned as keys and values.\vspace{.05in}
	\item 
	\textit{Feed-forward layer}: which finally generates the feedback as decoder's output.\vspace{.1in} 
\end{enumerate}



\subsubsection{Similarity Module}\label{sec:similarity} 
The similarity module uses a pre-trained Sentence Bidirectional Encoder Representations from Transformer (SBERT)~\cite{reimers2019sentence} based teacher-student model. The teacher and student models convert comments and feedback to language-agnostic vectors and produce their embeddings, $emb \textunderscore c$ and $emb \textunderscore f$. The language-agnostic vectors produce similar embeddings for the sentences using different words but portraying similar meaning. The teacher model $M$ maps the comments $c$ to $emb \textunderscore c$. The student model $\hat{M}$ maps the comments $c$ and feedbacks $f$ to $emb \textunderscore c$ and $emb \textunderscore f$, respectively. We have trained a student model $\hat{M}$ such that $\hat{M}$ ($emb \textunderscore c_i$) $\approx$ $M$($emb \textunderscore c_i$) and $\hat{M}$($emb \textunderscore f_i$) $\approx$ $M$($emb \textunderscore c_i$). The similarity score ($S_{score}$) between $emb \textunderscore c$ and $emb \textunderscore f$ is minimized as per Eq. \ref{eq:mse}. Taking MSE between them helps compute the closeness of their embeddings and hence the similarity between them. We have fine-tuned the similarity module using parallel comment-feedback pairs {(($c_1$, $f_1$), \dots ,($c_n$, $f_n$))} where the feedbacks with more than 80\% similarity during human evaluations have been considered.\vspace{-.1in} 

{
	\begin{eqnarray}
		\label{eq:mse}
		\begin{split}	
			&S_{score} = \frac{1}{n}\sum_{i \in n}^{} \Big( (\hat{M}(emb\_c_i) - M(emb\_c_i))^2 + (\hat{M}(emb\_f_i) - M(emb\_c_i))^2 \Big) \\
		\end{split}
	\end{eqnarray}
} 	 

\noindent Where $n$ is the number of feedback-comment pairs, whereas $emb\_c$ \& $emb\_f$ denote comment embedding and feedback embedding.

English SBERT model \cite{reimers2019sentence} has been used as the teacher model $M$ whereas RoBERTa \cite{liu2019roberta} has been used as the student model $\hat{M}$. The semantic similarities of the embeddings of the comments \& feedbacks have been found using cosine similarity as per Eq. \ref{eq:cossim}. \vspace{-.1in}

{
	\begin{eqnarray}
		\label{eq:cossim}
		\begin{split}	
			&\ \ \cos(emb\_c, emb\_f) = (emb\_c)^Temb\_f \\
		\end{split}
	\end{eqnarray}
} 	 
\noindent Where $emb\_c$ and $emb\_f$ are the embeddings for comment and feedback, respectively, and $T$ denotes the transpose operation.

To evaluate the feedbacks against ground-truth comments, Mean Reciprocal Rank (MRR) \& Recall@k (also known as `Recall Rate@k’) (defined in Section \ref{sec:metrics}). The MRR \& Recall@k are computed to denote whether the feedback generated is similar to the input comments, which are sorted based on the no. of likes. The `Recall@k' is computed for the generated feedback denoting whether it is similar to top $k$ ranked comments. The comments for an image are ranked based on their no. of likes. Then the semantic similarity between the comments' embeddings and the embeddings of generated feedback is found using cosine similarity. Finally, 'Recall@k' is computed for the generated feedback denoting whether it is similar to top $k$ ranked comments where $k$ is a user-definable integer. \\

\section{Experiments}\label{sec:experiments}	
\subsection{Training Strategy \& Parameter Tuning}\label{sec:training}
The baselines and proposed system described in the following sections have been trained for 30 epochs on GTX 1080Ti GPU with 3584 NVIDIA CUDA Cores and 11 GB GDDR5X Memory. The experiments have been performed using 5-fold cross-validation and 80\%-20\% training-testing split. The hyper-parameter values are described as follows.

\begin{itemize}
	\item 
	General parameters --
	batch-size: 32, learning rate: $5 \times 10^{-4}$ for text encoder \& $0.01$ for visual feature extraction model, network optimizer: Adam, loss function: cross entropy loss, activation function: ReLU. 
	
	\item 
	Parameters for the transformer model --
	input size: 512, output size: variable (as transformer model can generate the output with dynamic length), encoder embedding dimensions: 100, decoder embedding dimensions: 100, encoder hidden units dimensions: 128, decoder hidden units dimensions: 128, encoder dropout: 0.5, decoder dropout: 0.5, encoder no. of layers \& attention heads: 6 \& 8, decoder no. of layers \& attention heads: 6 \& 8, metric: accuracy. 

	\item 
	Parameters for the Faster R-CNN model --
	no. of epochs: 18, metric: mAP (mean Average Precision), no. of proposals: 36, no. of classes for anchor-boxes: 1601, network optimizer: adaDelta.
	
\end{itemize}

\subsection{Evaluation Metrics}\label{sec:metrics}
The generated feedbacks are evaluated using qualitative and quantitative metrics for their relevance with the ground-truth comments and inputs. We have incorporated two phases of automatic evaluation using five quantitative metrics (BLEU, CIDEr, ROUGE, SPICE, and METEOR) and two qualitative metrics (Recall@k and MRR) along with the human evaluation to evaluate the generated feedbacks holistically. For the quantitative evaluation of the feedbacks, the following metrics have been used in the first phase of automatic evaluation. These metrics can evaluate machine-generated sentences such as summaries, image descriptions, and translations against benchmark results and human references based on recall, precision, and sensitivity. 
\begin{itemize}
	\item \textbf{BLEU Score} \cite{papineni2002bleu}: BLEU (bilingual evaluation understudy) is a \textit{precision} based metric that compares candidate sentence with reference sentences to judge the quality of the candidate translation. In this work, 4-gram BLEU score has been used.
	\item \textbf{ROUGE} \cite{lin2004rouge}: ROUGE (Recall-Oriented Understudy for Gisting Evaluation) is a \textit{recall} based metric that analyzes automatic translation \& summaries with respect to ground-truth reference set. 
	\item \textbf{Meteor} \cite{lavie2009meteor}: METEOR (Metric for Evaluation of Translation with Explicit ORdering) considers \textit{precision} \& \textit{recall}'s harmonic mean to analyze automatically generated output at the sentence level. 
	\item \textbf{CIDEr} \cite{vedantam2015cider}: CIDEr (Consensus-based Image Description Evaluation) automatically evaluates machine translation and image caption outputs by considering the agreement of various reference descriptions.
	\item \textbf{SPICE} \cite{anderson2016spice}: SPICE (Semantic Propositional Image Caption Evaluation) is a metric based on the \textit{sensitivity} of the n-grams for automated evaluation of caption and sentences.
\end{itemize}

\noindent The second phase of automatic evaluation uses `Recall@k'~\cite{runeson2007detection} and `Mean Reciprocal Rank'~\cite{craswell2009mean} to judge the relevance of the generated feedbacks against the ground-truth comments. These metrics have been defined as follows.

\begin{itemize}
	\item \textbf{Mean Reciprocal Rank (MRR)}:
	If the $j^{th}$ feedback is most similar with the $k^{th}$ most liked comment, then the rank and reciprocal ranks of the $j^{th}$ feedback, i.e., $rank_j$ and $rrank_j$ are calculated as per Eq. \ref{eq:rr}.\vspace{-.1in} 
	
	\begin{eqnarray}
		\label{eq:rr}
		\begin{split}		
			&rank_j = k\\
			&rrank_j = 1/k\\
		\end{split}
	\end{eqnarray}	
	
	\noindent Where $k$ denotes the $k^{th}$ comment sorted by the number of likes whereas $rank_j$ \& $rrank_j$ are the Rank \& Reciprocal rank of the $j^{th}$ feedback.
	
	Mean Reciprocal Rank (MRR) denotes the average of the reciprocal ranks of all the feedback samples considered in the study, and it is given by Eq. \ref{eq:mrr}.\vspace{-.1in} 
	
	\begin{eqnarray}
		\label{eq:mrr}
		\begin{split}		
			&MRR = (\frac{1}{Q})\sum_{j=1}^{Q} \frac{1}{rank_j}\\
		\end{split}
	\end{eqnarray}
	
	\noindent Where $Q$ is the number of feedback samples and $rank_j$ denotes the rank of the $j^{th}$ feedback.\\
	
	\item \textbf{Recall@k}:
	In general, Recall@k denotes whether a data sample matches with any of the top $k$ relevant samples. It has been adapted to evaluate whether the generated feedback matches with top $k$ relevant ground-truth comments. We have found the similarity score of the feedback and find the ranks for all the comments where rank denotes the rank of the corresponding comment among all the comments when sorted by no. of likes. As shown in Eq. \ref{eq:recall}, if the comment with which feedback shows maximum similarity score is in top $k$ comments, it will get a score of 1 for Recall@k, else 0. 
	
	\begin{eqnarray}
		\label{eq:recall}
		\begin{split}		
			&Recall@k = 1\ \text{if}\ rank_q \in [1, \dots, k]\\ 
		\end{split}
	\end{eqnarray}
	
	\noindent Where $q$ is the $q^{th}$ feedback; $k$ is the $k^{th}$ comment sorted by number of likes and $rank_j$ denotes the rank of the $j^{th}$ feedback.
	
\end{itemize}

\noindent Moreover, to ensure that the generated feedbacks represent and are relevant to the given inputs, they have been evaluated by the human readers alongside the objective measures. Manual evaluation has also been carried out by having 50 human evaluators read the generated feedbacks and evaluate them against input image, text, and comments for their similarity. 

\subsection{Ablation Studies}\label{sec:ablation}
Ablation studies have been performed to analyze the effect of using visual information, attention mechanism, and region extraction. Table \ref{tab:performance1} summarises these studies.

\subsubsection{Effect of using Visual Modality}\label{sec:ablatev}
The feedbacks are first generated considering only the text input, and then the complementary visual information has also been considered. The qualitative scores improved on including visual modality along with the textual modality. 

\subsubsection{Effect of using Attention}
The model is first trained on textual attention only; then, the visual attention is also fed to further fine-tune the output. It has been observed that the output with both textual and visual attention was more human-like than the one with textual attention only. It should be noted that the scores reduced a bit on including visual features along with textual, though they improved significantly on including the attention mechanism. 

\subsubsection{Effect of using Region Proposal}
Better performance has been observed on incorporating region extraction along with using textual \& visual information with attention. In Section \ref{sec:ablatev} global visual features using VGG and ResNet were included, whereas this Section incorporates local features using Faster R-CNN based region proposals. The aforementioned ablation studies on using visual modality, attention, and region proposal have been summarized in Table \ref{tab:performance1}. 

\begin{table}[H]
	\centering
	{\fontsize{8}{11}\selectfont
		\caption{Ablation studies on using textual \& visual modalities, attention and region proposal. Here, `T', `V', `A' and `R' denote Textual \& Visual features, Attention \& Region extraction.}
		\label{tab:performance1}
		\resizebox{.6\textwidth}{!}{
			\begin{tabular}{@{}l|lllll@{}}
				\toprule
				\textbf{Architecture}       & \textbf{BLEU} & \textbf{CIDEr}& \textbf{ROUGE}& \textbf{SPICE} & \textbf{METEOR} \\ \midrule
				{T Only}                    &0.2137	&0.1365	&0.2667	&0.1282	&0.1192\\
				{T + V}                     &0.2073	&0.1482	&0.2486	&0.1344	&0.1132\\
				{T + V + A}                 &0.3046	&0.1884	&0.3515	&0.1783	&0.1422\\
				{T + V + A + R}             &0.3023	&0.1945	&0.3842	&0.1792	&0.1638\\ \bottomrule
			\end{tabular}
	}} 
\end{table} 

\subsubsection{Effect of using Data Samples with Varying Ranges of Comments}
The experiments are performed using complete data, data with low (up to 5), mid (between 13 and 50), and high (more than 30) no. of comments per image. The split was done using the following thresholds -- low-comments: up to 5 comments per image or tweet, mid-comments: between 13 and 50 comments per image, and high-comments: more than 30 comments per image. The performance for various combinations is shown in Table \ref{tab:performance2}. It is observed that experimenting with the complete dataset produced similar results as samples with 13 to 50 comments; however, it was computationally expensive. 

\begin{table}[H]
	\centering
	{\fontsize{8}{11}\selectfont
		\caption{Ablation studies on using data samples with various ranges of comments.}
		\label{tab:performance2}
		\resizebox{.68\textwidth}{!}{
			\begin{tabular}{@{}l|lllll@{}}
				\toprule
				\textbf{Comments per image}       & \textbf{BLEU} & \textbf{CIDEr}& \textbf{ROUGE}& \textbf{SPICE} & \textbf{METEOR} \\ \midrule
				{Complete Data} 		    &0.3023	&0.1945	&0.3842	&0.1792	&0.1638\\ 
				{Low-comments (upto 5)}     &0.1734	&0.1204	&0.2263	&0.1456	&0.1345\\ 
				{Mid-comments (13-50)}      &0.2992	&0.2082	&0.3717	&0.1768	&0.1559\\ 
				{High-comments (30+)}       &0.2656	&0.1737	&0.3218	&0.1362	&0.1235\\ \bottomrule
			\end{tabular}
	}} 
\end{table} 

\subsubsection{Computational Time Analysis}
Table \ref{tab:time} shows the time taken by various configurations to train the model for 1 epoch. Though including region extraction is computationally expensive, it resulted in best performance than other configurations (Observed from Table \ref{tab:performance1}). On the other hand, using 13-50 comments per image (mid-comments) resulted in similar quantitative performance scores as training with complete data (Observed from Table \ref{tab:performance2}). The final implementation has been carried out using region proposal along with visual \& textual attention and the data samples with 13-50 comments per image.

\begin{table}[H]
	\centering
	{\fontsize{8}{11}\selectfont
		\caption{Computation time (in hours) to train 1 epoch of the model for various configurations.}
		\label{tab:time}
		\resizebox{.48\textwidth}{!}
		{
			\begin{tabular}{@{}l|cc@{}}
				\toprule
				\textbf{Architecture}  & \textbf{Complete Data} & \textbf{Mid-comments} \\ \midrule
				T Only        & 17.01 h         & 5.18 h        \\
				T + V         & 20.04 h        & 6.23 h        \\
				T + V + A     & 1.78 h        & 0.75 h     \\
				T + V + A + R & 137 h          & 47.5 h         \\ \bottomrule
			\end{tabular}
		}
	}
\end{table}

\subsection{Models}\label{sec:models}
The architectures of the baselines and proposed system have been determined based on the ablation studies performed in Section \ref{sec:ablation}. As observed earlier, the feedbacks generated considering textual, and corresponding visual features had better scores; hence, all the baseline models have included textual and visual encoders. 
\begin{itemize} 
	\item \textbf{Baseline 1}: The first baseline uses Gated Recurrent Units (GRU) for the textual encoder, whereas residual network (ResNet) is used for the visual encoder network. The choice of GRU over Long Short Term Memory (LSTM) architecture is guided through experimental observations. LSTM corresponded to 20\% more memory consumption as compared to GRU, whereas their performance was comparable.\vspace{.05in}
	
	\item \textbf{Baseline 2} -- The second baseline retains ResNet for visual encoding; however, it replaces the textual encoder network with Bidirectional GRU (BiGRU). The intuition behind using BiGRU was to enable a particular word's embedding to embody its contextual meeting, which changes according to the words appearing before and after it. \vspace{.05in} 
	
	\item \textbf{Baseline 3} -- Textual Encoder: This model further replaces the textual encoder with a text transformer model, whereas the attention mechanism is incorporated along with the ResNet-based visual encoder. The choice of using text transformer is governed by its off-the-shelf performance in various language analysis problems such as summarization, translation, caption generation, etc.~\cite{vaswani2017attention}.\vspace{.05in}
	
	\item \textbf{Proposed System} -- Textual Encoder: The proposed system uses text transformer as textual encoder, whereas a Faster R-CNN based region proposal mechanism is incorporated along with ResNet extracted features for visual encoding. The incorporation of local features extracted by R-CNN resulted in a better performance as compared to using only the global features extracted by ResNet.\vspace{.1in}
\end{itemize} 
The proposed system's implementation code and MMFeed dataset constructed in this paper can be accessed at \href{https://github.com/MIntelligence-Group/MMFeed}
{\underline {github.com/MIntelligence-Group/MMFeed.}}\vspace{.15in}

\section{Results and Discussion}\label{sec:results}
The feedbacks generated by the proposed system have been evaluated using the quantitative and qualitative measures described in Section~\ref{sec:metrics}, and they have been compared with the Baseline methods. 

\subsection{Quantitative Results}
In the first phase of automatic evaluation, the generated feedbacks are quantitatively evaluated using BLEU, CIDEr, ROUGE, SPICE, and METEOR metrics. The results have been shown in Table~\ref{tab:quantre}, which indicate the increased informative-ness for the feedbacks generated by the proposed method as compared to the baseline models.

\begin{table}[!h]
	\centering
	{\fontsize{8}{11}\selectfont
		\caption{Quantitative evaluation of the generated feedbacks.}
		\label{tab:quantre}
		\resizebox{.6\textwidth}{!}{
			\begin{tabular}{@{}l|ccccl@{}}
				\toprule
				\textbf{Model}& \textbf{BLEU}& \textbf{CIDEr}& \textbf{ROUGE}& \textbf{SPICE}& \textbf{METEOR} \\ \midrule
				\textbf{Baseline 1}    &0.1942	&0.1342	&0.2524	&0.1025 &0.0924\\
				\textbf{Baseline 2}    &0.2124	&0.1735	&0.2745	&0.1654	&0.1393\\
				\textbf{Baseline 3}    &\textbf{0.3096}	&0.1835	&0.3374	&0.1554	&0.1412\\ \midrule[.01pt]
				\textbf{Proposed}      &0.3023	&\textbf{0.1945}	&\textbf{0.3842}	&\textbf{0.1792}	&\textbf{0.1638}\\ \bottomrule
			\end{tabular}
	}} 
\end{table}	

\begin{figure}[!b]
	\vspace{-.1in}
	\centering
	\captionsetup{justification=centering}
	\subfloat[Sample Result 1]
	{\includegraphics[width=0.85\textwidth]{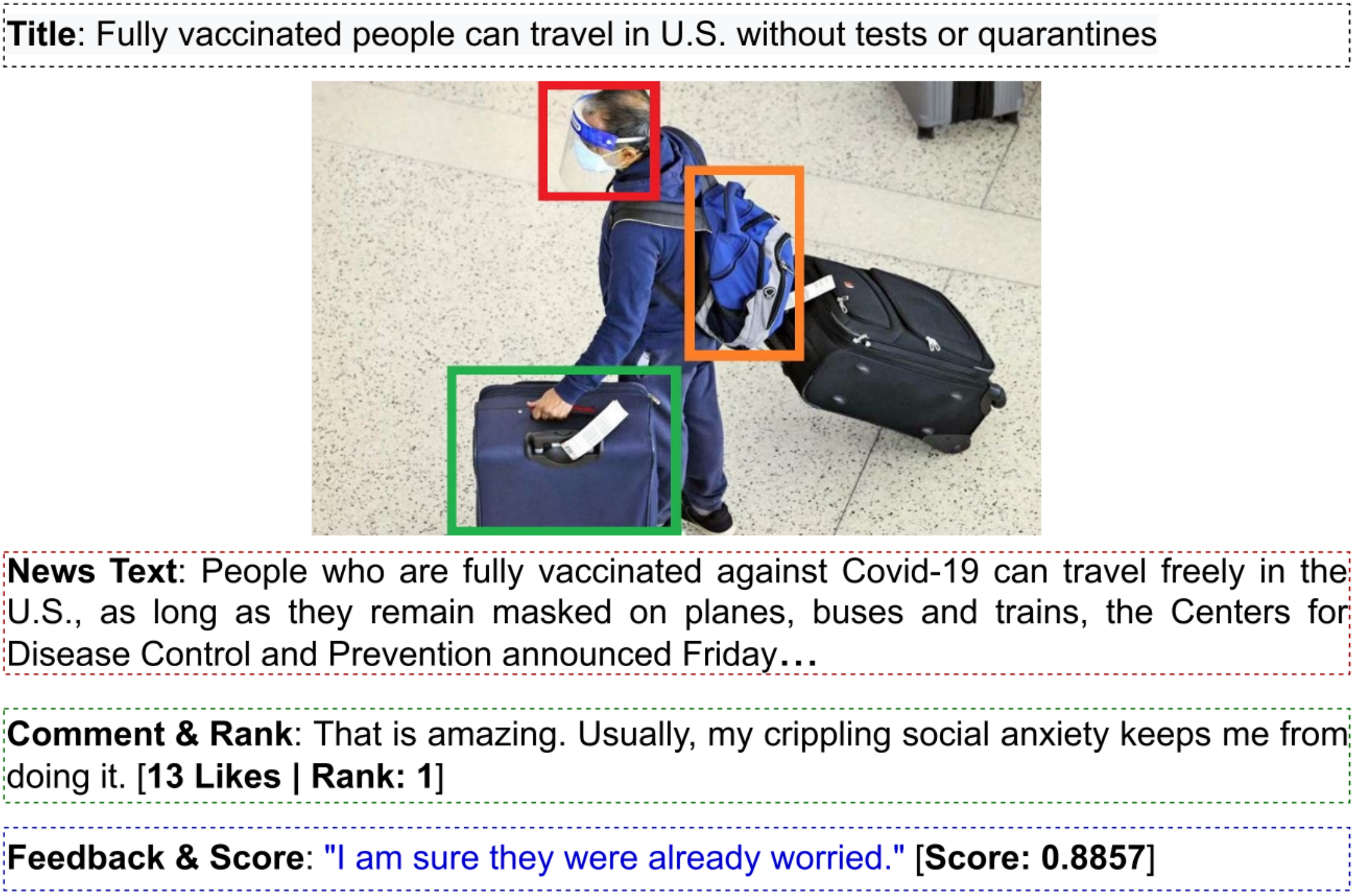}
		\label{fig:f1}}
	
	\vspace{.15in}
	\subfloat[Sample Result 2]
	{\includegraphics[width=0.85\textwidth]{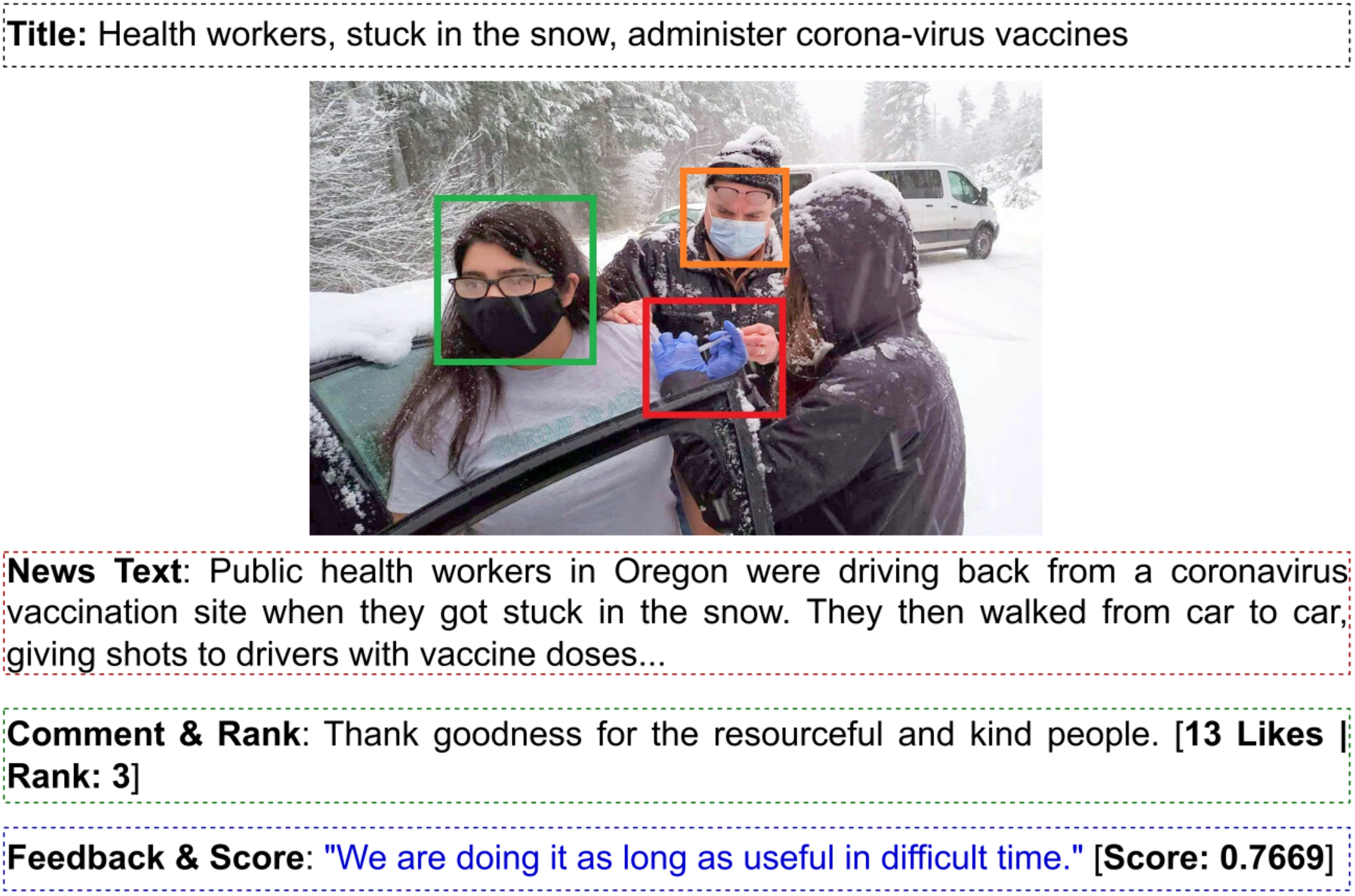}
		\label{fig:f2}}
	\addtocounter{figure}{1} 
\end{figure}
\begin{figure}[!t]	
	\ContinuedFloat
	\centering
	\captionsetup{justification=centering}
	\subfloat[Sample Result 3]
	{\includegraphics[width=0.85\textwidth]{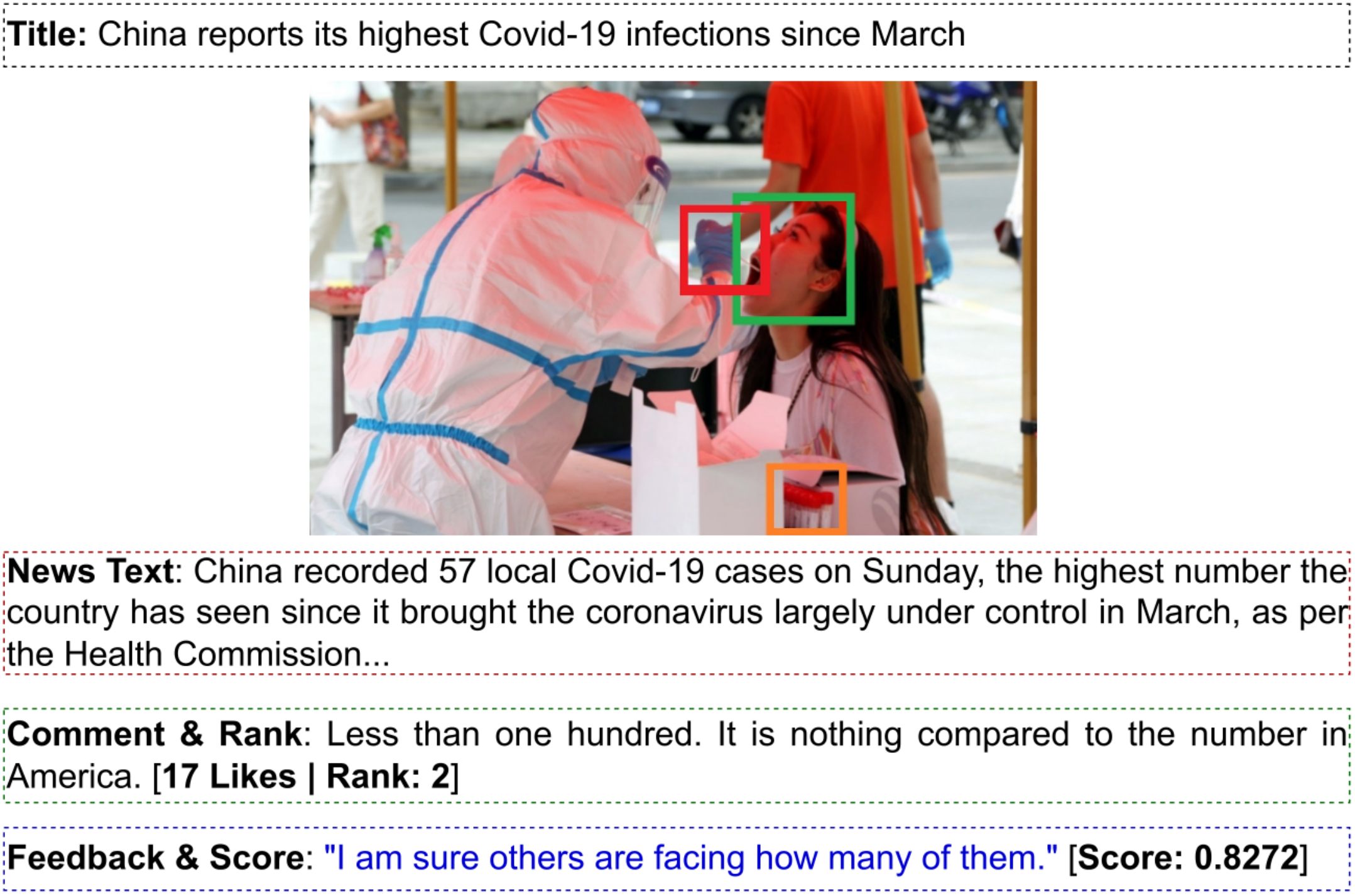}
		\label{fig:f2}}
	
	\vspace{.15in}
	\subfloat[Sample Result 4]
	{\includegraphics[width=0.85\textwidth]{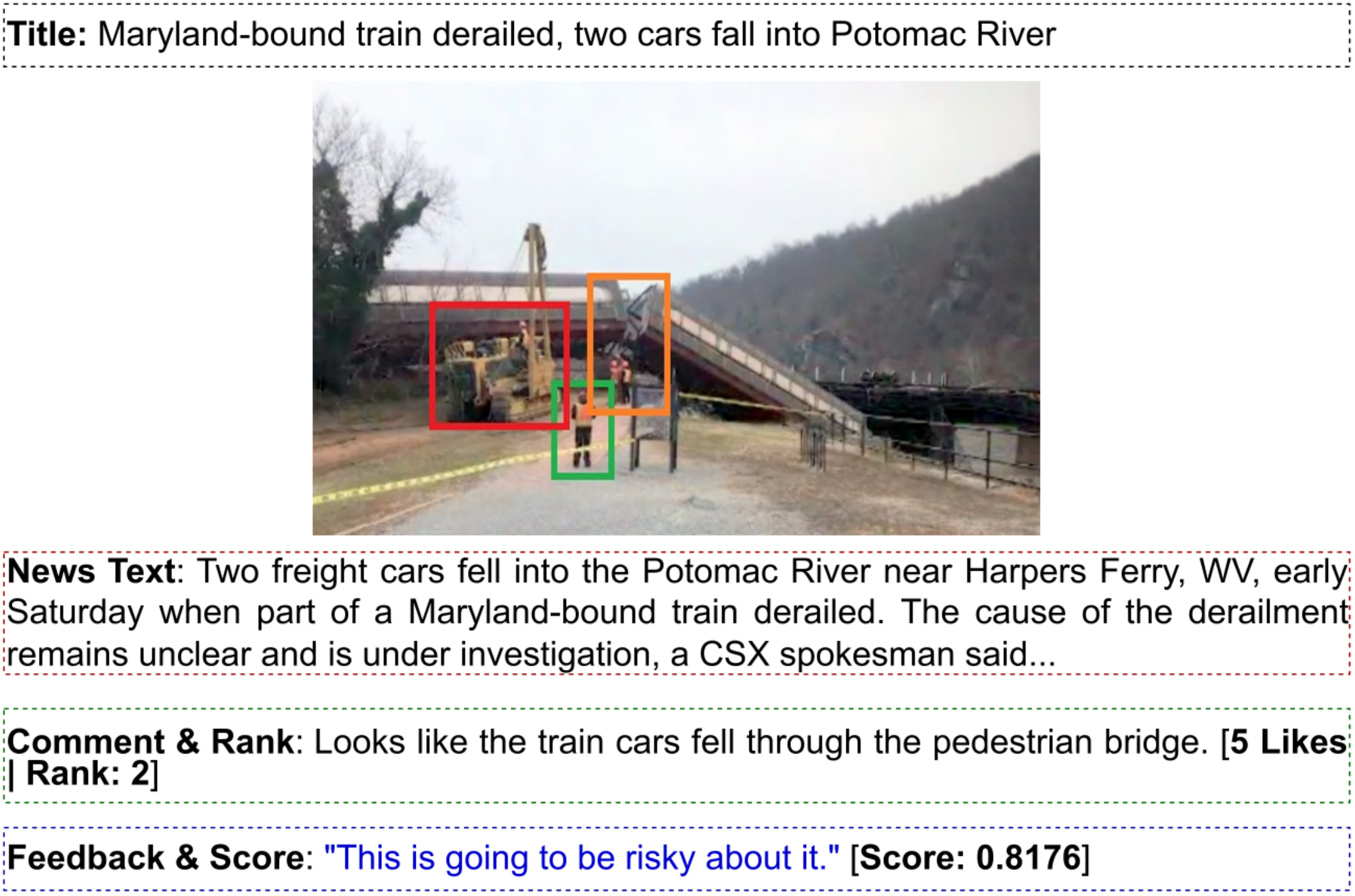}
		\label{fig:f3}} 
	\caption{Sample Results. Here, `Score' denotes the highest of the similarity scores of the predicted feedback with all the comments, and `Rank' denotes the rank of the corresponding comment among all the comments when sorted by no. of likes.}
	\label{fig:res_Qual_a}
\end{figure}

\subsection{Qualitative Results}\label{sec:res_Qual_a}
The second automatic evaluation phase evaluates the generated feedbacks using MRR and Recall@k metrics. As discussed in Table~\ref{tab:qualres}, 98.67\% of the feedbacks are relevant to one of the top 7 ground-truth comments, whereas they show the MRR of 0.3042, denoting that a majority of the feedbacks portrayed the most similarity with the top 3 (corresponding to MRR of 0.33) or top 4 (corresponding to MRR of 0.25) comments itself. 

\begin{table}[H]
	\centering
	{\fontsize{8}{11}\selectfont
		\caption{Automatic evaluation. Here, `MMR' \& `R@k' denote ‘Mean Reciprocal Rank' \& `Recall@k.'}
		\label{tab:qualres}
		\resizebox{.55\textwidth}{!}{
			\begin{tabular}{@{}l|ccccc@{}}
				\toprule
				\textbf{Model}& \textbf{MRR}& \textbf{R@1}& \textbf{R@3}& \textbf{R@5}& \textbf{R@7} \\ \midrule
				\textbf{Baseline 1}    &0.2412		&24.36		&70.56		&91.52		&93.63 \\
				\textbf{Baseline 2}    &0.2643		&25.23		&71.42		&93.53		&96.32 \\
				\textbf{Baseline 3}    &0.2923	 	&26.42	 	&79.97		&\textbf{98.76}		&95.92 \\ \midrule[.01pt]
				\textbf{Proposed}      &\textbf{0.3042}	 	&\textbf{29.33}	 	&\textbf{84.56}		&98.32		&\textbf{98.67} \\ \bottomrule
			\end{tabular}
	}} 
\end{table}

The ground-truth comments and generated feedback portray similar sentiment and context using different words. To ensure that the generated feedbacks represent and are relevant to the given inputs, they have also been evaluated by the human readers alongside the objective measures. Table \ref{tab:quantres} summarises the results of the same. It should be noted that the `Score' described in Fig.~\ref{fig:res_Qual_a} is for the automatic evaluations of the generated feedbacks while `$s$' described in Table \ref{tab:quantres} corresponds to their human evaluation.

\begin{table}[H]
	\centering
	{\fontsize{8}{11}\selectfont
		\caption{Human evaluation. Here, `$s_{ct}$', `$s_{ci}$', `$s_{ft}$', `$s_{fi}$', and `$s_{cf}$' denote semantic similarities between comments \& text, comment \& image, feedbacks \& text, feedbacks \& image, and comments \& feedbacks as judged by human readers.}
		\label{tab:quantres}
		\resizebox{.55\textwidth}{!}{
			\begin{tabular}{@{}l|ccccl@{}}
				\toprule
				\textbf{Model}& $\mathbf{s_{ct}}$& $\mathbf{s_{ci}}$& $\mathbf{s_{ft}}$& $\mathbf{s_{fi}}$& $\mathbf{s_{cf}}$ \\ \midrule
				\textbf{Baseline 1}    &47.90	&41.12	&43.40	&36.07	&45.12	\\
				\textbf{Baseline 2}    &49.33	&45.93	&46.33	&41.67	&58.22	\\
				\textbf{Baseline 3}    &64.11	&64.33	&58.78	&59.67	&71.44	\\ \midrule[.01pt]
				\textbf{Proposed}      &\textbf{72.86}	&\textbf{74.90}	&\textbf{65.34}	&\textbf{67.96}	&\textbf{80.17}	\\\bottomrule
			\end{tabular}
	}} 
\end{table}


As per the human evaluations, the ground-truth comments show 72.86\% and 74.90\% similarity with the input text and image, respectively. These numbers are 65.34\% and 67.96\% respectively for the generated feedbacks. At the same time, the similarity between the feedbacks and comments has been found to be 80.17\%.

\subsection{Discussion}\label{sec:discussion}
The proposed system has been trained using human comments along with input text \& images to generate feedback towards multimodal content just as humans do. The evaluation results advocate that the generated feedbacks are relevant to the corresponding text \& image input. We hope that the novel problem of multimodal human-like feedback synthesis and the MMFeed dataset proposed in this paper will inspire the researchers for further advancements in this context. 

The proposed system's architecture has been evolved through progressive experiments. The model is first trained on textual attention only; then, the visual attention is also fed to fine-tune the output further. The output with both textual \& visual attention has been observed to be more human-like than the one with textual attention only. Further ablation studies have been performed to analyze the effect of the feature extraction techniques. Various architecture choices for textual and visual encoders have also been evaluated. Finally, a transformer-based textual encoder and Faster R-CNN-based visual encoder have been implemented. The incorporation of the region proposal increased the computational cost to 63 times because of the large size of the bounding boxes. On the other hand, using the data samples with 13-50 comments per image for training resulted in similar performance as training with complete data, although it took 5.5X less computational time. Finally, a trade-off was chosen between training the proposed system with region proposals and using the data samples with 13-50 comments per image. Further, the baselines and proposed system converged in terms of validation loss in 18-23 epochs. The models have been trained for 30 epochs as a safe upper bound.

The generated feedbacks have been evaluated for relevance with the input image \& text and ground-truth comments using two automatic and one manual evaluation phases. The improvements in the evaluation scores on incorporating attention \& region proposal advocate that the generated feedbacks represent the corresponding text \& image input. The automatic \& human evaluation results affirm that the generated feedbacks are relevant to the input text \& image. The generated feedbacks have been observed to learn the in-context information from the training data. For example, many training data samples contained information about politics and corona-virus. The context about the same was reflected in some of the generated feedback. 


One challenge with the evaluation of affective feedback generation is that multiple feedback can be contextually similar and may convey the same information. Thus given the ground truth, the evaluation is difficult as our goal is to produce contextually similar feedback. The meaning of the term `Human-like' is to be taken more in the sense that the model can generate contexts similar to how a human would do. Though the syntax \& semantics of the feedbacks are not entirely correct, the proposed system can learn and generate the words with respect to the text and image inputs. The minor errors in the model's understanding of the human language can be attributed to the noise (special characters, sentence phrases, and multilingual symbols) present in the comments of the training data.

\section{Conclusions and future work}\label{sec:conclusion}	
In this work, we have introduced a novel task to generate human-like feedback for text \& image data and proposed an affective feedback synthesis system. We have also constructed a large-scale multimodal feedback synthesis dataset. Automatic and human evaluations have been carried out to evaluate the generated feedbacks' relevance with the input text \& image and similarity with the comments. 

In the future, we aim to improve the syntactic and semantic correctness while generating long feedback sentences and extend the dataset to include more than one image per news article. It is also planned to work on the news articles' sentiment classification and genre classification. For the automatic evaluation of the generated feedbacks, existing evaluation metrics such as SPICE, CIDEr, ROUGE, etc., have been used, which are broadly used for various types of machine-generated sentences such as automatic translations, summaries, and image captions. There is a need to design an automatic evaluation metric specifically for evaluating multimodal feedbacks that would aid or replace the human evaluation process. We will focus on that as well in our future work. 


\section*{Acknowledgements}
Ministry of Education INDIA has supported this research through grant reference no. 1-3146198040.

\appendix
\section{Appendix: Data Crawling \& Pre-processing Procedure}\label{appendix1}
Algorithm \ref{algo:a1} shows the procedure to crawl and pre-process the data instances during the construction of the MMFeed dataset.\vspace{.1in}


{\small
	\begin{algorithm}[H]
			\SetAlgoLined
			\textbf{Define} $user\_name$: Twitter handle of the news channel\\ 
			\textbf{Define} $tweet\_id$: A unique numeric ID of particular Tweet\\ 
			\textbf{Define} $tweet$: Tweet contents\\ 
			\textbf{Define} $URL$: URL link of the original news article\\ 
			\textbf{Define} $text$: Text fetched from the original news article\\ 
			\textbf{Define} $image$: Image fetched from the original news article\\ 
			\textbf{Define} $replies$: Comments fetched from Tweet contents\\
			\textbf{Define} $replies\_iter$: Iteration, denoting the count of replies\\ \vspace{.07in}
			
			procedure crawl(curr\_url): \hfill $\triangleright$ {\color{gray}Definition of the crawler function}\\
			\hspace{.1in} article = Article(curr\_url) \hfill $\triangleright$ {\color{gray}`curr\_url' denotes the current url}\\
			\hspace{.1in} article.download() \hfill $\triangleright$ {\color{gray}Function to download the article contents}\\
			\hspace{.1in} article.parse() \hfill $\triangleright$ {\color{gray}Function to parse the article contents}\\\vspace{.07in}
			
			\hspace{.1in} for full\_tweets in tweepy.Cursor() \hfill $\triangleright$ {\color{gray}To crawl Tweet text using Tweepy library's Cursor function}\\ 
			\hspace{.1in} \hspace{.1in} tweet=full\_tweets.full\_text \hfill $\triangleright$ {\color{gray}Tweet Contents}\\
			\hspace{.1in} \hspace{.1in} replies\_iter = tweepy.Cursor() \hfill $\triangleright$ {\color{gray}Comments}\\\vspace{.07in} 
			
			\hspace{.1in} While True\\
			\hspace{.1in} \hspace{.1in} reply=replies\_iter.next()\\
			\hspace{.1in} \hspace{.1in} replies.append(reply.full\_text)\\
			\hspace{.1in} \hspace{.1in} favorite.append(reply.\_json['favorite\_count']) \hfill $\triangleright$ {\color{gray}No. of likes}\\\vspace{.07in}
			
			
			news\_text = crawl(curr\_url)\hfill $\triangleright$ {\color{gray}Calling the crawl procedure to fetch the contents of current url}\\\vspace{.07in}
			
			with open (str(iter)+'twitter\_data.csv','w', encoding='utf-8') as csv\_file: \hfill $\triangleright$ {\color{gray}Save to a CSV file}\\
			\hspace{.1in}	writer = csv.DictWriter(csv\_file, fieldnames = [Tweet, Comment, Likes])\\
			\hspace{.1in}   writer.writerow({'Tweet': tweet, 'Comment': ':'.join(replies), 'Likes': ':'.join(favorite)})\\\vspace{.1in}
			
			$\triangleright$ {\color{gray}Data pre-processing}\\	
			procedure normalised\_text(text):\\
			\hspace{.1in} Strip html tags (text)  \hfill $\triangleright$ {\color{gray}Parse using `BeautifulSoup' library and remove HTML tags}\\
			\hspace{.1in} Expand contractions (text)\hfill $\triangleright$ {\color{gray}For example, don't $\rightarrow$ do\ not}\\\vspace{.07in}
			
			data=read\_csv(twitter\_data.csv) \hfill $\triangleright$ {\color{gray}Load the saved CSV file after data crawling}\\
			\hspace{.1in} title = data[`Tweet'][0] \hfill $\triangleright$ {\color{gray}News Title}\\
			\hspace{.1in} text = article.text \hfill $\triangleright$ {\color{gray}Input text}\\
			\hspace{.1in} text = normalised\_text(text) \\
			\hspace{.1in} image = urllib.request.urlopen(article.top\_image) \hfill $\triangleright$ {\color{gray}Input image}\\
			\hspace{.1in} replies=data[`Comment'][0]\\
			\hspace{.1in} replies = normalised\_text(replies) \\
			\hspace{.1in} likes=data[`Likes'][0]\\\vspace{.07in}
			
			$\triangleright$ {\color{gray}Save pre-processed data to `mmfeed\_data.csv'\ file}\\
			with open (str(iter)+'mmfeed\_data.csv','w', encoding='utf-8') as csv\_file:\\
			\hspace{.1in}	writer = csv.DictWriter(csv\_file, fieldnames=[`Title', `Text', `Image', `Comment', `Likes'])\\
			\hspace{.1in}   writer.writerow({`Title': title, `Text': text, `Image': image, `Comment': replies, `Likes': favorite})\\\vspace{.07in}   
			
			\caption{MMFeed Dataset Crawling \& Pre-processing Procedure}
			\label{algo:a1} 
	\end{algorithm}
}
\vspace{.1in}
\noindent The MMFeed dataset along with the proposed system's implementation code can be accessed through this link: \href{https://github.com/MIntelligence-Group/MMFeed}
{\underline {github.com/MIntelligence-Group/MMFeed.}} 



\bibliographystyle{ACM-Reference-Format}
\bibliography{MMFeed}

\end{document}